\documentclass[proof]{pasj00}

\begin{document}
\SetRunningHead{S.Tanaka , K.Yoshikawa , T.Okamoto and K.Hasegawa}{A new ray-tracing implementation for 3D diffuse radiation transfer}

\title{A new ray-tracing scheme for 3D diffuse radiation transfer 
on highly parallel architectures}

%

%
 \author{
   Satoshi \textsc{Tanaka}\altaffilmark{1}
   Kohji \textsc{Yoshikawa}\altaffilmark{1}
   Takashi \textsc{Okamoto}\altaffilmark{2}
   and
   Kenji \textsc{Hasegawa}\altaffilmark{1,3}
}
   \altaffiltext{1}{Center for Computational Sciences, University of Tsukuba, 1-1-1 Tennodai, Tsukuba Ibaraki 305-8577}
   \email{stanaka@ccs.tsukuba.ac.jp}
   \email{kohji@ccs.tsukuba.ac.jp}
   \altaffiltext{2}{Department of Cosmosciences, Hokkaido University, Kita 10 Nishi 8, Kita-ku, Sapporo Hokkaido 060-0810}
   \email{okamoto@astro1.sci.hokudai.ac.jp}
   \altaffiltext{3}{Graduate School of Science, Nagoya University,
   Furo-cho, Chikusa-ku, Nagoya Aichi 464-8602}
   \email{hasegawa.kenji@a.mbox.nagoya-u.ac.jp}


\maketitle

\begin{abstract}
We present a new numerical scheme to solve the transfer of diffuse
radiation on three-dimensional mesh grids which is efficient on
processors with highly parallel architecture such as recently popular
GPUs and CPUs with multi- and many-core architectures. The scheme is
based on the ray-tracing method and the computational cost is
proportional to $N_{\rm m}^{5/3}$ where $N_{\rm m}$ is the number of
mesh grids, and is devised to compute the radiation transfer along each
light-ray completely in parallel with appropriate grouping of the
light-rays. We find that the performance of our scheme scales well with
the number of adopted CPU cores and GPUs, and also that our scheme is
nicely parallelized on a multi-node system by adopting the multiple wave
front scheme, and the performance scales well with the amount of the
computational resources. As numerical tests to validate our scheme and
to give a physical criterion for the angular resolution of our
ray-tracing scheme, we perform several numerical simulations of the
photo-ionization of neutral hydrogen gas by ionizing radiation sources
without the ``on-the-spot'' approximation, in which the transfer of
diffuse radiation by radiative recombination is incorporated in a
self-consistent manner.
\end{abstract}

\section{Introduction}

Radiation transfer (RT) has been long recognized as a indispensable
ingredient in numerically simulating many astrophysical phenomena
including the reionization of intergalactic medium (IGM) in the early
universe, radiative feedback during the galaxy formation, and others. So
far, varieties of numerical schemes for solving the RT in three
dimensions are proposed during the last two decades \citep{iliev2006},
and some of them can be coupled with the hydrodynamic simulations
\citep{iliev2009} thanks to not only the increase of the available
computational resources, but also the improvement of numerical
algorithms to solve the RT in many astrophysical conditions.

Most of the numerical schemes for RT can be divided into two groups: one
is the moment-based schemes which solve the moment equation of the RT
equation instead of solving the RT equation directly, and the other is
the ray-tracing schemes. As for the moment-based schemes, the important
advantage is that the computational costs scale with the number of mesh
grids, $N_{\rm m}$ and hence can be easily coupled with hydrodynamic
simulations. The flux-limited diffusion (FLD) scheme, which adopts the
closure relation valid in the diffusion limit, is the most common among
the moment-based schemes, while there are a number of more sophisticated
schemes which close the moment equations with the optically thin
variable Eddington tensor approximation \citep{gnedin2001} and the
locally evaluated Eddington tensor (the M$_1$ model)
\citep{gonzalez2007, skinner2013, kanno2013}.  The accuracy and validity
of the moment-based schemes are, however, problem-dependent. For
example, the FLD scheme has a problem in handling shadows formed behind
opaque objects \citep{gonzalez2007}. While schemes with M$_1$ model are
capable of simulating shadows sucessfully, they cannot solve the
crossing of multiple beamed lights, where the beamed lights unphysically
merge into one beam \citep{rosdahl2013}. Therefore, the ray-tracing
schemes are naturally chosen for solving the RT in situations that we
are considering in the studies of galaxy formation and cosmic
reionization, in which there exist a number of radiation sources.

In ray-tracing schemes, emission and absorption of radiation are
followed along the light-rays that extend through the computational
domain. As for the long-characteristics schemes
\citep{abel1999,sokasian2001} in which light-rays between all radiation
sources and all other relevant meshes are considered, the computational
cost scales with $N_{\rm m}^{2}$ in general cases and $N_{\rm
m}^{4/3}N_{\rm s}$ when we consider only the RT from point radiating
sources, where $N_{\rm s}$ is the number of point sources. On the other
hand, for the short-characteristics schemes \citep{kunasz1988,stone1992}
which are similar to the long-characteristics schemes but integrate the
RT equation only along paths connecting nearby mesh grids, the
computational cost scales with $N_{\rm m}^{5/3}$ in general and $N_{\rm
m}N_{\rm s}$ for the RT from point sources. Ray-tracing schemes are in
principle versatile for any physical settings but computationally much
more expensive than the moment-based schemes. Due to such huge
computational costs, RT simulations with the ray-tracing schemes have
been applied only to static conditions or snapshots of hydrodynamical
simulations in a post-process manner in many previous studies.

Some of the ray-tracing schemes are now coupled with hydrodynamical
simulations adopting smoothed particle hydrodynamics (SPH) codes
\citep{susa2006, hasegawa2010, pawlik2011} and mesh-based codes
\citep{rijkhorst2006, wise2011}, and they can handle the RT and its
hydrodynamical feedback in a self-consistent manner. Majority of these
radiation hydrodynamics codes, however, consider the transfer of
radiation only from point sources and ignore the effect of radiation
transfer from spatially extended diffuse sources, such as the
recombination radiation emitted from ionized regions and infrared
radiation emitted by dust grains, since the computational costs for
computing the transfer of diffuse radiation is prohibitively large.

Specifically, in the numerical RT calculations of the hydrogen ionizing
radiation, we usually adopt the on-the-spot approximation in which one
assumes that the ionizing photons emitted by radiative recombinations in
ionized regions are absorbed by neutral atoms in the immediate vicinity
of the recombining atoms. However, adopting the on-the-spot
approximation can fail to notice the important effects of diffuse
recombination radiation in some situations.  The roles of ionizing
recombination photons in the epoch of cosmic reionization is discussed
by a number of works \citep{ciardi2001, miralda-escude2003, dopita2011,
rahmati2013a}. \citet{dopita2011} proposed the recombination photons
produced in the fast accretion shocks in the structure formation in the
universe as an possible source of ionizing photons responsible for the
cosmic reionization, though \citet{wyithe2011} showed that its impact on
the cosmic reionization is not very significant. It is also reported
that the recombination radiation plays an important role at transition
regions between highly ionized and self-shielded regions
\citep{rahmati2013a}. As for the effect of recombination photons on the
galaxy-size scales, \citet{inoue2010} showed that the recombination
radiation produces the Lyman-`bump' feature in the spectral energy
distributions of high-$z$ galaxies, and also that the escaping ionizing
photons from high-$z$ galaxies are to some extent contributed by the
recombination radiation.  \citet{rahmati2013b} also pointed out that the
recombination radiation makes the major contribution to the
photo-ionization at regions where the gas is self-shielded from the UV
background radiation.

The RT of infrared diffuse radiation emitted by dust grains plays an
important roles in the evolution of star-forming galaxies, in which the
radiation pressure exerted by multi-scattered infrared photons drives
stellar winds. In most of numerical simulations of galaxy formation,
however, such momentum transfer is treated only in a phenomenological
manner (e.g. Okamoto et al. 2014).

In this paper, we present a new ray-tracing scheme to solve the RT of
diffuse radiation from spatially extended radiating sources efficiently
on processors with highly parallel architectures such as graphics
processing units (GPUs) and multi-core CPUs which are recently popular
or available in near future. The basic idea of the scheme is based on
the scheme presented by \citet{razoumov2005} and `Authentic Radiation
Transfer' (ART) scheme \citet{nakamoto2001b}.  Generally speaking,
development of such numerical schemes with high concurrency is of
critical importance because the performance improvement of recent
processors are achieved by the increase of the number of processing
elements or CPU cores integrated on a single processor chip rather than
the improvement of the performance of individual processing elements.

The rest of the paper is organized as follows. Section 2 is devoted to
describe the numerical scheme to simulate the radiation transfer. In
section 3, we present our implementation of the scheme suitable to
highly parallel architectures such as GPUs and CPUs with multi-core
architectures. We present the results of numerical test suits of RT of
diffuse radiation in Section 4. The computational performance of our
implementation is shown in Section 5. Finally, we summarize our results
in Section 6.

\section{Methodology}

In this section, we describe our ray-tracing scheme of diffuse radiation
transfer. Generally, the radiation field can be decomposed into two
components. One is the direct incident radiation from point radiation
sources, and the other is the diffuse radiation emerged from spatially
extended regions. In our implementation, the RT of photons emitted by
point radiation sources is computed separately from that of diffuse
radiation. Throughout in this paper, we consider the RT of hydrogen
ionizing photons emitted by point radiation sources, and recombination
photons emerged from the ionized regions as the diffuse radiation. We
use the steady state RT equation for a given frequency $\nu$:
\begin{equation}
\frac{dI_{\nu}}{d\tau_{\nu}} = -I_{\nu} +  \mathcal{S}_{\nu} ,
\label{eq:rad_tr1}
\end{equation}
where $I_{\nu},\tau_{\nu}$ and $\mathcal{S}_{\nu}$ are the specific
intensity, the optical depth and the source function, respectively.  The
source function is given by $\mathcal{S}_{\nu} =
\varepsilon_{\nu}/\kappa_{\nu},$ where $\kappa_{\nu}$ and
$\varepsilon_{\nu}$ are the absorption and emission coefficients,
respectively.  The formal solution of this equation is given by
\begin{equation}
I_{\nu}(\tau_{\nu}) = I_{\nu}(0)\,e^{-\tau_{\nu}} + \int^{\tau_{\nu}}_{0} \mathcal{S}_{\nu}(\tau'_{\nu}) e^{-(\tau_{\nu}-\tau'_{\nu})}d\tau'_{\nu} ,
\label{eq:rad_tr2}
\end{equation}
where $\tau'_{\nu}$ is the optical depth at a position along the ray.
When we adopt the ``on-the-spot'' approximation in which recombination
photons emitted in ionized regions are assumed to be absorbed where they
are emitted, we neglect the source function, $\mathcal{S}_\nu$, and the
formal solution is simply reduced to
\begin{equation}
I_{\nu}(\tau_{\nu}) = I_{\nu}(0)\,e^{-\tau_{\nu}} .
\label{eq:on_the_spot}
\end{equation}

\subsection{RT from point radiation sources\label{sub:rt_point}}

\begin{figure}
 \centering \includegraphics[width=5cm]{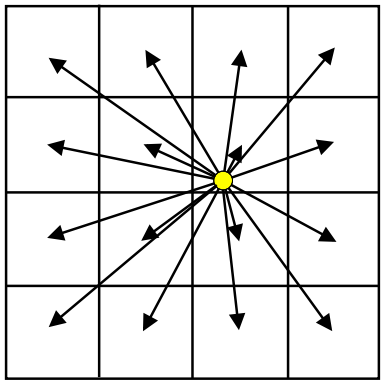}
 \caption{Schematic illustration of the ray-tracing method for the
 radiation emitted by a point radiation source in the
 two-dimensional mesh grids.\label{fig:long}}
\end{figure}

To solve the RT from point radiation sources, we compute the optical
depth between each pair of a point radiation source and a target mesh
grid, i.e. an end point of each light-ray (see
Figure~\ref{fig:long}). Instead of solving equation
(\ref{eq:on_the_spot}), we compute the radiation flux density at the
target mesh grid as
\begin{equation}
 f_\alpha(\nu) = \frac{L_\alpha(\nu)}{4\pi r_\alpha^2}\,\exp\left[-\tau_\alpha(\nu)\right],
\end{equation}
where $L_\alpha(\nu)$ is the intrinsic luminosity of the $\alpha$-th
point radiation source, and $r_\alpha$ and $\tau_\alpha(\nu)$ are the
distance and the optical depth between the point radiation source and the
target mesh grid, respectively. Then, the photo-ionization and
photo-heating rates of the $i$-th species contributed by the $\alpha$-th
point radiation source are computed by
\begin{equation}
 \Gamma^\alpha_{i,\gamma} = \int^\infty_{\nu_i}
  \frac{f_\alpha(\nu)}{h\nu}\sigma_i(\nu)\,d\nu,
\end{equation}
and 
\begin{equation}
 \mathcal{H}^\alpha_{i,\gamma} = \int^\infty_{\nu_i} \frac{f_\alpha(\nu)}{h\nu}(h\nu-h\nu_i)\sigma_i(\nu)\,d\nu
\end{equation}
respectively, where $\sigma_i(\nu)$ and $\nu_i$ are the ionization cross
section and the threshold frequency of the $i$-th species,
respectively. In the test simulations desribed in this paper, we compute
these photo-ionization and photo-heating rates in a photon-conserving
manner \citep{abel1999} as described in appendix
~\ref{app:photon_conserving}.

For a single point radiation source, the number of rays to be calculated
is $N_{\rm m}$, and the number of mesh grids traveled by a single
light-ray is in the order of $N^{1/3}_{\rm m}$. Thus, the computational
cost for a single point radiation source is proportional to
$N^{4/3}_{\rm m}$. Therefore, the total computational cost scales as
$N_{\rm m}^{4/3} N_{\rm s}$, where $N_{\rm s}$ is the number of point
radiation sources. For a large number of point radiation sources, we can
mitigate the computational costs by adopting more sophisticated scheme
such as the ARGOT scheme \citep{okamoto2012} in which a distant group of
point radiation sources is treated as a bright point source located at
the luminosity center with a luminosity summed up for all the sources in
the group to effectively reduce the number of radiation sources and
hence the computational cost is proportional to $N^{4/3}_{\rm m} \log
N_{\rm s}$.

\subsection{RT of the diffuse radiation \label{sub:rt_rec}}
\begin{figure}
 \centering \includegraphics[width=5cm,clip]{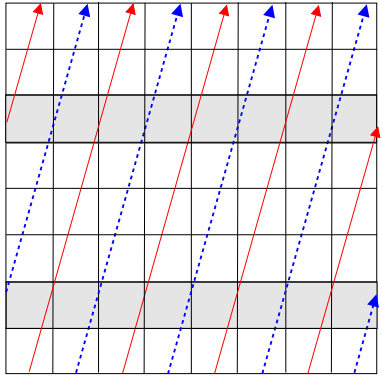}
 \caption{Schematic illustration of the ray-tracing scheme for the
 diffuse radiation in the two-dimensional mesh grid. For a given
 direction, equally-spaced parallel light-rays are cast from boundaries
 of the simulation volume and travel to the other boundaries. Note that
 gray mesh grids are traversed by multiple parallel light-rays, while
 the subsets of light-rays depicted by blue or red get through them only
 once.  }  \label{fig:parallel_ray}
\end{figure}

We solve the equation~(\ref{eq:rad_tr2}) to compute the RT of the
diffuse radiation. The numerical scheme we adopt in this work is based
on the method developed by \citet{razoumov2005} and ART scheme
\citep{nakamoto2001b, iliev2006}, which is reported to have little
numerical diffusion in the searchlight beam test like the long
characteristics method although its computational cost is proportional to
$N_{\rm m}^{5/3}$ similarly to that of the short characteristic method
\citep{nakamoto2001b}. In this scheme, we solve the
equation~(\ref{eq:rad_tr2}) along equally spaced parallel rays as
schematically shown in Figure~\ref{fig:parallel_ray}.

For a given incoming radiation intensity $I^{\rm in}_\nu$ along a
direction $\hat{\mathbf{n}}$, the outgoing radiation intensity $I^{\rm
out}_\nu$ after getting through a path length $\Delta L$ of a single
mesh is computed by integrating equation~(\ref{eq:rad_tr2}) as
\begin{equation}
 I^{\rm out}_\nu (\hat{\mathbf{n}}) = I^{\rm in}_\nu(\hat{\mathbf{n}})\,e^{-\Delta\tau_\nu} + \mathcal{S}_\nu
  (1-e^{-\Delta\tau_\nu}), 
  \label{eq:rt}
\end{equation}
where $\Delta\tau_\nu$ is the optical depth of the path length $\Delta
L$ (i.e. $\Delta\tau_\nu = \kappa_\nu \Delta L$), and $\mathcal{S}_\nu$
and $\kappa_\nu$ are the source function and the absorption coefficient
of the mesh grid, respectively.

The intensity of the incoming radiation averaged over the path length
$\Delta L$ across a single mesh grid can be calculated as
\begin{equation}
 \bar{I}^{\rm in}_\nu(\hat{\mathbf{n}}) = \frac{1}{\Delta L}\int_0^{\Delta L} I^{\rm
  in}_\nu (\hat{\mathbf{n}})e^{-\kappa_\nu l}\,dl = I_\nu^{\rm in}(\hat{\mathbf{n}})\frac{1-e^{-\Delta
  \tau_\nu}}{\Delta \tau_\nu}.
\end{equation}
In addition to this, we have a contribution to the radiation intensity
from the source function which we set constant in each mesh grid, and
the total intensity averaged over the path length is given by
\begin{equation}
 \bar{I}_\nu(\hat{\mathbf{n}}) = \bar{I}^{\rm in}_\nu(\hat{\mathbf{n}}) + \mathcal{S}_\nu = I_\nu^{\rm in}(\hat{\mathbf{n}}_i)\frac{1-e^{-\Delta
  \tau_\nu}}{\Delta \tau_\nu} + \mathcal{S}_\nu
\end{equation}
For those mesh grids through which multiple parallel light-rays pass,
the averaged intensity can be given by
\begin{equation}
 \bar{I}_\nu^{\rm ave}(\hat{\mathbf{n}}) = 
  \frac{\sum_j \Delta \tau_{\nu,j} \bar{I}_{\nu,j}(\hat{\mathbf{n}})}{\sum_j \Delta
  \tau_{\nu,j}} 
  = \bar{I}_{\nu}^{\rm ave, in}(\hat{\mathbf{n}}) + \mathcal{S}_\nu, 
  \label{eq:path_averaged_intensity}
\end{equation}
where $\bar{I}_{\nu,i}$ and $\Delta \tau_{\nu, i}$ are the intensity averaged
over the $i$-th light-ray and the optical depth of $i$-th light-ray in
the mesh grids, respectively, $\bar{I}^{\rm ave, in}_\nu$ is a
contribution from the incoming radiation given by
\begin{equation}
 \bar{I}^{\rm ave, in}_\nu(\hat{\mathbf{n}}) = \frac{\sum_j \Delta \tau_{\nu,j}
  \bar{I}^{\rm in}_{\nu,j}(\hat{\mathbf{n}})}{\sum_j \Delta
  \tau_{\nu,j}},
 \label{eq:averaged_intensity}
\end{equation}
and the summation is over all the parallel light-rays in the same mesh
grid.  Then, the mean intensity can be computed by averaging
$\bar{I}^{\rm ave}_\nu$ described above over all the directions as,
\begin{equation}
 J_\nu = \frac{1}{N_{\rm d}}\sum_{i=1}^{N_{\rm d}} \bar{I}_{\nu}^{\rm
  ave}(\hat{\mathbf{n}}_i) = J^{\rm in}_\nu + \mathcal{S}_\nu,
  \label{eq:mean_intensity}
\end{equation}
where $\hat{\mathbf{n}}_i$ describes a vector toward the $i$-th
direction and $N_{\rm d}$ is the number of directions of light-rays to
be considered, $\bar{I}_{\nu}^{\rm ave}(\hat{\mathbf{n}}_i)$ is the
averaged intensity along the $i$-th direction calculated with
equation~(\ref{eq:path_averaged_intensity}), and $J^{\rm in}_\nu$ is
given by
\begin{equation}
J^{\rm in}_\nu=\frac{1}{N_{\rm d}}\sum_{i=1}^{N_{\rm d}} \bar{I}^{\rm
 ave, in}_{\nu}(\hat{\mathbf{n}}_i).
\end{equation}
Then, the photo-ionization and photo-heating rates of the $i$-th species
contributed by the diffuse radiation in each mesh grid can be computed
as
\begin{equation}
 \Gamma_{i,\gamma}^{\rm diff} = 4\pi\int_{\nu_i}^\infty \frac{J_\nu}{h\nu}\sigma_i(\nu)\,d\nu
\end{equation}
and 
\begin{equation}
 \mathcal{H}_i^{\rm diff} = 4\pi \int_{\nu_i}^\infty
  \frac{J_\nu}{h\nu}(h\nu-h\nu_i) \sigma_i(\nu)\,d\nu
\end{equation}

As for the recombination radiation of ionized hydrogen (HII) regions,
the number of recombination photons to the ground state per unit time
per unit volume, $\dot{N}^{\rm rec}$, can be expressed in terms of the
emissivity coefficient $\varepsilon_\nu$ as
\begin{equation}
 \dot{N}^{\rm rec}=4\pi \int_{\nu_0}^{\infty}\frac{\varepsilon_\nu}{h\nu}\,d\nu =
  [\alpha_{\rm A}(T)-\alpha_{\rm B}(T)]n_{\rm e}n_{\rm HII},
\end{equation}
where $\nu_0$ is the Lyman limit frequency, $\alpha_{\rm A}(T)$ and
$\alpha_{\rm B}(T)$ are the recombination rates of HII as functions of
temperature $T$ in the case-A and case-B approximations, respectively,
and $n_{\rm e}$ and $n_{\rm HII}$ are the number densities of the
electrons and HII, respectively.
In this work, we adopt a rectangular functional form of
$\varepsilon_\nu/(h\nu)$ as
\begin{equation}
 \frac{\varepsilon_\nu}{h\nu} = \left\{
                    \begin{array}{ll}
                     \displaystyle \frac{\Delta\alpha(T)n_{\rm
                      e}n_{\rm HII}}{4\pi \Delta\nu_{\rm th}} &
                      (\nu_0\le\nu\le\nu_0+\Delta\nu_{\rm th}) \\
                     0 & (\mbox{otherwise}),
                    \end{array}
                   \right.
\end{equation}
where $\Delta\alpha(T)=\alpha_{\rm A}(T)-\alpha_{\rm B}(T)$ and
$\Delta\nu_{\rm th}$ is the frequency width of the recombination
radiation and given by $\Delta\nu_{\rm th} = k_{\rm B}T/h$. Thus, the
source function is given by
\begin{equation}
 \mathcal{S}_\nu = \frac{\varepsilon_\nu}{\kappa_\nu}  =
  \left\{
   \begin{array}{ll}
    \displaystyle \frac{\Delta\alpha(T)n_{\rm e}n_{\rm HII}h\nu}{4\pi n_{\rm
     HI}\sigma_{\rm HI}(\nu) \Delta\nu_{\rm th}} &
     (\nu_0\le\nu\le\nu_0+\Delta\nu_{\rm th} )\\
    0& ({\rm otherwise}).
   \end{array}
  \right.
\end{equation}
This spectral shape is the same as adopted in \citet{kitayama2004} and
\citet{hasegawa2010}, the results of which are compared with our results
to verify the validity of our scheme. Note that for the typical
temperature of HII regions, $T=10^4\,\,{\rm K}$, we have $\Delta\nu_{\rm
th}\ll \nu_0$.  Calculations of the mean intensity based on
equations~(\ref{eq:rt}) to (\ref{eq:mean_intensity}) are done in a
monochromatic manner at the Lyman limit frequency $\nu_0$.  In computing
photo-ionization and photo-heating rates, we assume that the mean
radiation intensity $J^{\rm in}_\nu$ has a rectangular functional form
as
\begin{equation}
 J^{\rm in}_\nu = \left\{
                    \begin{array}{ll}
                     \displaystyle \mathcal{J}^{\rm in} &
		      (\nu_0\le\nu\le\nu_0+\Delta\nu_{\rm th}) \\
                     0 & (\mbox{otherwise}).
                    \end{array}
                   \right.
\end{equation}
Therefore, the photo-ionization and photo-heating rates of neutral
hydrogen can be rewritten as
\begin{equation}
 \Gamma^{\rm diff}_{\rm HI,\gamma} = 4\pi\mathcal{J}^{\rm
  in}\int_{\nu_0}^{\nu_0+\Delta\nu_{\rm th}}\frac{\sigma_{\rm HI}(\nu)}{h\nu} d\nu + \frac{\Delta\alpha(T)n_{\rm
  e}n_{\rm HII}}{n_{\rm HI}},
  \label{eq:gamma_diff}
\end{equation}
and 
\begin{equation}
 \mathcal{H}^{\rm diff}_{\rm HI,\gamma} = 4\pi\mathcal{J}^{\rm
  in}\int_{\nu_0}^{\nu_0+\Delta\nu_{\rm th}}\left(1-\frac{\nu_0}{\nu}\right)\,\sigma_{\rm HI}(\nu)\,d\nu +
  \frac{\Delta\alpha(T)n_{\rm e}n_{\rm HII}}{2n_{\rm HI}}h\Delta\nu_{\rm
  th},
  \label{eq:heat_diff}
\end{equation}
respectively. In the test simulations described in
section~\ref{sec:test}, we fix the frequency width $\Delta\nu_{\rm th}$
by assuming temperature of HII regions to be $10^4$ K, and the integrals
in equations~(\ref{eq:gamma_diff}) and (\ref{eq:heat_diff}) can be
estimated prior to the simulations. For the transfer of diffuse
radiation with more general spectral form, we can easily extend our
method described above by adopting a nonparametric functional form of
radiation spectra as
\begin{equation}
 I_{\nu}=\sum_i \mathcal{I}^i \Pi(\nu-\nu_i,\Delta\nu),
\end{equation}
where $\Pi(x,y)$ is the rectangular function given by
\begin{equation}
 \Pi(x,y) = \left\{\begin{array}{ll}
	     1 & -y/2 \le x \le y/2\\
		    0 & {\rm otherwise} \\
		   \end{array}\right.,
\end{equation}
and $\nu_i$ is the central frequency of the $i$-th frequency bin. 

The number of light-rays parallel to a specific direction is
proportional to $N_{\rm m}^{2/3}$, and the number of mesh grids
traversed by a single light-ray is in the order of $N_{\rm
m}^{1/3}$. Therefore, the total computational cost is proportional to
$N_{\rm m}N_{\rm d}$.

\subsection{Angular resolution for RT of the diffuse radiation} 
\label{sub:direction} 

The number of the directions of light-rays, $N_{\rm d}$, determines
angular resolution of the RT of the diffuse radiation. In order to
guarantee that light-rays from a mesh grid on a face of the simulation
box reach all the mesh grids on the other faces, $N_{\rm d}$ should be
in the order of $N_{\rm m}^{2/3}$. In the case that the mean free path
of the diffuse photons is sufficiently shorter the simulation box size,
however, such a large $N_{\rm d}$ is redundant because only a small
fraction of diffuse photons reach the other faces, and we can reduce the
total computational cost by decreasing the number of directions, $N_{\rm
d}$, while keeping the reasonable accuracy of the diffuse RT. Thus, the
number of directions should be flexibly changed depending on the
physical state.

To achieve this, we use the HEALPix (Hierarchical Equal Area isoLatitude
Pixelization) software package \citep{gorski2005} to set up the
directions of the light-rays. The HEALPix is suitable to our purposes in
the sense that each direction corresponds to exactly the same solid angle
and that the directions are nearly uniformly sampled. Furthermore, it
can provide a set of directions with these properties in arbitrary
resolutions, each of which contains $12N_{\rm side}^2$ directions, where
$N_{\rm side}$ is an angular resolution parameter. Since it is larger
than the number of mesh grids on six faces of a cube with a side length
of $N_{\rm side}$ mesh spacings, $6N_{\rm side}^2$, it is expected that
a set of light-rays originated from a single point with directions
generated by the HEALPix with an angular resolution parameter of $N_{\rm
side}$ get through all the mesh grids within a cube centered by the
point with a side length of $N_{\rm side}$ mesh spacings. Thus, the
optimal number of directions should be chosen so that the mean free path
of the recombination photons is sufficiently shorter than $N_{\rm
side}\Delta H$, where $\Delta H$ is the mesh spacing.

\subsection{Chemical reactions and radiative heating and cooling}

With photo-ionization and photo-heating rates computed with the
prescription described above, time evolutions of chemical compositions
and thermal states of gas are computed in the same manner as adopted in
\citet{okamoto2012}. Details of the numerical schemes are briefly
described in appendices~\ref{app:reaction},\ref{app:heating_cooling} and
\ref{app:timestep}.

The chemical reaction rates and radiative cooling rates adopted in this
paper are identical to those adopted in \citet{okamoto2012}, and the
literatures from which we adopt these rates are summarized in Table
\ref{tab:rate}.

\begin{table} 
\caption{Rates of chemical reactions and radiative cooling processes
 adopted in this paper. Reference for radiative recombination rates (RR)
 of HII, HeII and HeIII in the case-A and case-B approximation;
 collisional ionization rates (CIR) of HI, HeI, and HeII; recombination
 cooling rates (RCR) of HII, HeII and HeIII in the case-A and case-B
 approximation; collisional ionization cooling rates (CICR) of HI, HeI
 and HeII; collisional excitation cooling rates (CECR) of HI, HeI and
 HeII; bremsstrahlung cooling rate; inverse Compton cooling rate (CCR);
 photoionization cross sections (CS) of HI, HeI and HeII. \label{tab:rate}}
 \centering
 \begin{tabular}{ll}
  \hline
  physical process & literature \\
  \hline
  RR (case-A) & (1), (1), (2) \\
  RR (case-B) & (3), (3), (3) \\
  CIR & (7), (7), (1) \\
  RCR (case-A) & (2), (2), (2) \\
  RCR (case-B) & (3), (5), (3) \\
  CICR & (2), (2), (2) \\
  CECR & (2), (2), (2) \\
  BCR & (4) \\
  CCR & (6) \\
  CS & (8), (8), (8)\\
  \hline
 \end{tabular}
 \begin{flushleft}
  (1) \citet{abel1997}; 
  (2) \citet{cen1992}; 
  (3) \citet{hui1997}; 
  (4) \citet{hummer1994};
  (5) \citet{hummer1998};
  (6) \citet{ikeuchi1986};
  (7) \citet{janev1987}; 
  (8) \citet{agnagn}; 
 \end{flushleft}
\end{table}

\section{Implementation on Highly Parallel Architectures}

In this section, we describe the details of the implementation of the RT
calculation of the diffuse radiation which performs effectively on
recently popular processors with highly parallel architecture, such as
GPUs, multi-core CPUs, and many-core processors. Throughout this paper,
we present the results using the implementation with the
\verb|OpenMP| and \verb|CUDA| technologies. The former is supported by
most of the multi-core processors, and the many-core processors such as
the Intel Xeon-Phi processor, while the latter is the parallel
programming platform for GPUs by NVIDIA.

In the implementation on GPUs with the \verb|CUDA| platform, the fluid
dynamical and chemical data in all the mesh grids are transferred from
the memory attached to CPUs to those of GPUs prior to the RT
calculations. After the RT calculations, ionization rates and heating
rates in all the mesh grids computed on GPUs are sent back to the CPU
memory. 

\subsection{Ray Grouping}\label{ss:ray_grounping}

In the calculations of the transfer of the diffuse radiation described
in the previous section, many parallel light-rays travel from boundaries
of the simulation volume until they reach the other boundaries. On
processors with highly parallel architecture, a straightforward
implementation is to assign a single thread to compute the RT along each
light-ray and calculate the RT along multiple light-rays in
parallel. Such a simple implementation, however, does not work because
some mesh grids are traversed by multiple parallel light-rays (see gray
mesh grids in Figure~\ref{fig:parallel_ray}), and in computing
equation~(\ref{eq:path_averaged_intensity}), multiple computational
threads write data to the identical memory addresses. Thus,
equation~(\ref{eq:path_averaged_intensity}) has to be computed not in
parallel but in a exclusive manner using the ``atomic operations''.  The
use of the atomic operations, however, significantly degrade the
parallel efficiency and computational performance in many architectures.

\begin{figure}
 \centering \includegraphics[width=6.9cm]{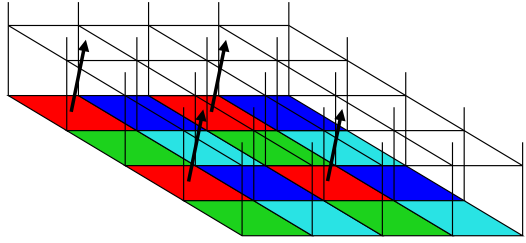}
 \caption{Schematic illustration of light-ray grouping for the
 three-dimensional mesh grids.  Light-rays in each group start from
 boundary faces of mesh grids painted with the same color. Only the
 light-rays in one group are shown in this figure. \label{fig:grouping}}

\end{figure}

To avoid such use of atomic operations and the deterioration of the
parallel efficiency, we split the parallel light-rays into several
groups so that parallel light-rays in each group do not traverse any
mesh grid more than once. For example, in two-dimensional mesh grids in
Figure~\ref{fig:parallel_ray}, parallel light-rays are split into two
groups each of which are depicted by blue and red arrows. One can see
that light-rays in each groups do not intersect any mesh grids more than
once. We can extend this technique to the three-dimensional mesh grids
by splitting the parallel light-rays into four groups, where the
light-rays in each group are cast from the two-dimensionally interleaved
mesh grids as depicted by the same color in Figure~\ref{fig:grouping}.

\subsection{Efficient Use of Multiple External Accelerators}
Many recent supercomputers are equipped with multiple external
accelerators such as GPUs on a single computational node, each of which
has an independent memory space. To attain the maximum benefit of the
multiple accelerators, we decompose calculations of the diffuse
radiation transfer according to the directions of the light-rays, and
assign the decomposed RT calculation to the multiple accelerators. After
carrying out the RT calculation for the assigned set of directions, and
computing the mean intensity with equation (\ref{eq:mean_intensity})
averaged over the partial set of directions on each external
accelerator, the results are transferred to the memory on the hosting
nodes. Then, we obtain the mean intensity averaged over all directions.

\subsection{Node Parallelization}
In addition to the thread parallelization within processors, we
implement the inter-node parallelization using the Message Passing
Interface (MPI). In the inter-node parallelization, the simulation box
is evenly decomposed into smaller rectangular blocks with equal volumes
along the Cartesian coordinate.

For the inter-node parallelization of the calculations of the diffuse
radiation transfer, we adopt the multiple wave front (MWF) scheme
developed by \citet{nakamoto2001}, in which light-ray directions are
classified into eight groups according to signs of their three direction
cosines, and for each group of light-ray directions, the RT calculations
along each direction are carried out in parallel on a ``wave front'' in
the node space, while the RT for different directions are computed on
the other wave fronts simultaneously. By transferring the radiation
intensities at the boundaries from upstream nodes to downstream ones,
one can sequentially compute the RT of diffuse radiation along all the
directions in each group of light-ray directions. See
\citet{nakamoto2001} for more detailed description of MWF scheme.

\section{Test Simulation}
\label{sec:test}

In this section, we present a series of test simulations to validate our
RT code. All the test simulations are carried out with $128^3$ mesh
grids and the angular resolution parameter of $N_{\rm side}=8$ unless
otherwise stated.

\subsection{Test-1 : HII region expansion}
\begin{figure}
 \centering 
 \includegraphics[width=15cm]{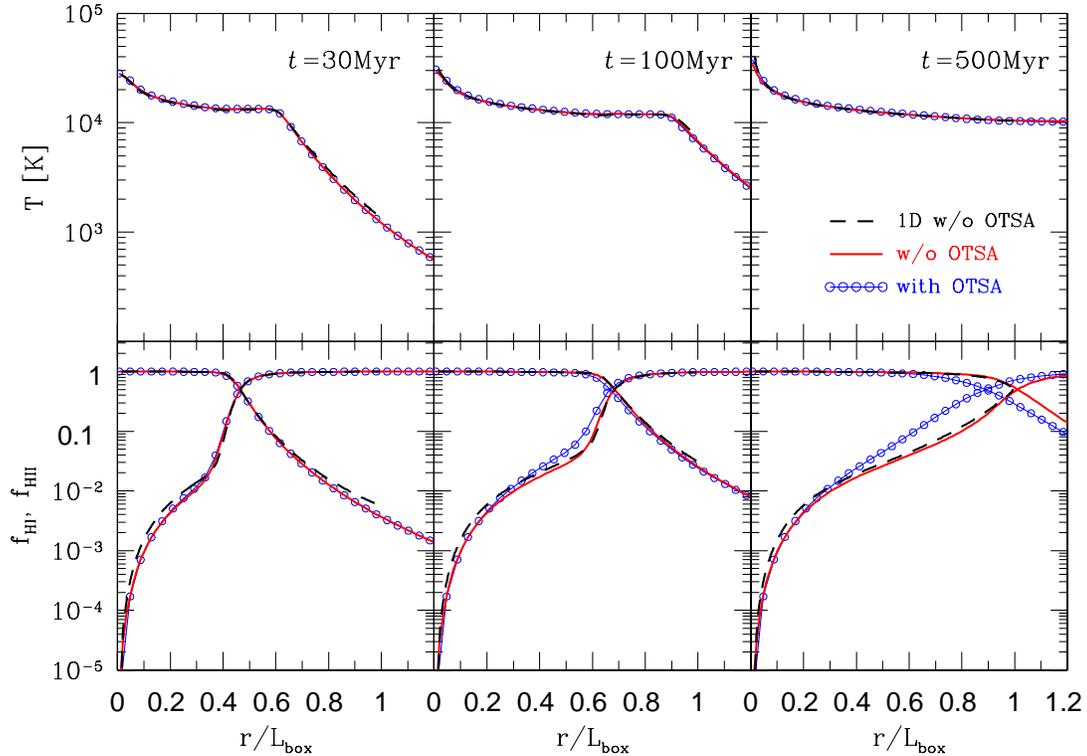}

 \caption{Test-1: Radial profiles of neutral and ionized fractions of
 hydrogen and gas temperature at $t=30{\rm Myr}$, $100{\rm Myr}$ and
 $500{\rm Myr}$. Solid lines with and without circles indicates the
 results with and without the on-the-spot approximation (OTSA),
 respectively.  Dashed lines show the results obtained with
 one-dimensional spherically symmetric code without the OTSA presented
 in \citet{kitayama2004}. \label{fig:test-1}}
\end{figure}

The first test is the simple problem of a HII region expansion in a
static homogeneous gas which consists of only hydrogen around a single
ionizing source.  We adopt the same initial condition as that of Test-2
in Cosmological Radiative Transfer Codes Comparison Project I
\citep{iliev2006}, where the hydrogen number density is $n_{\rm
H}=10^{-3}\,{\rm cm}^{-3}$ and the initial gas temperature is
$T=100\,{\rm K}$. The ionizing source emits the blackbody radiation with
an effective temperature of $10^5\,{\rm K}$, and $5\times 10^{48}$
ionizing photons per second and located at a corner of simulation box
with a side length of 6.6 kpc. In this initial condition, the
recombination time is $t_{\rm rec}=122.4{\rm\, Myr}$ and the
Str\"{o}mgren radius is estimated to be 5.4 kpc.
Figure~\ref{fig:test-1} shows the radial profiles of ionization fraction
and gas temperature at $t=30{\rm Myr}$, $100{\rm Myr}$ and $500{\rm
Myr}$. The solid lines with and without circles indicate the results
with and without the on-the-spot approximation (OTSA) , respectively.
In the calculation with the effect of recombination radiation, the
ionized regions are more extended than those computed with the
on-the-spot approximation , especially at later stages ($t=100{\rm Myr}$
and $500{\rm Myr}$) because of the additional ionization of hydrogen by
the recombination photons.

To verify the validity of our scheme for the transfer of diffuse
recombination radiation, we compare our results with the ones obtained
with the one-dimensional spherically symmetric RT code by
\citet{kitayama2004}, which also incorporates the transfer of
recombination photons emitted by the ionized hydrogen using the
impact-parameter method. We find that the one-dimensional results with
the effect of recombination radiation denoted by dashed lines show a
good agreement with our three-dimensional results, indicating that our
treatment of diffuse radiation transfer is consistent with that of
well-established impact-parameter method.

\subsection{Test-2 : Shadow by a dense clump} 
\label{sub:angular_resolution} 

In the second test, we compute the RT from point radiation source in the
presence of a dense gas clump. A point radiation source is located at
the center of the simulation box with the same side length as the Test-1
(6.6kpc), and surrounded by the ambient uniform gas with the same
hydrogen number density and temperature as the Test-1 ($n_{\rm
H}=10^{-3}\,{\rm cm}^{-3}$ and $T=100\,{\rm K}$, respectively). In
addition, we set up a spherical dense gas clump with a radius of
0.56\,kpc centered at 0.8 kpc apart from the point radiation source
along the $x$-direction. We set the density of the dense clump to 200
times higher than that of the ambient gas, and the temperature is set to
100\,K. The spectrum and luminosity of the point radiation source is the
same as that in Test-1.

\begin{figure}
 \centering \includegraphics[width=10cm]{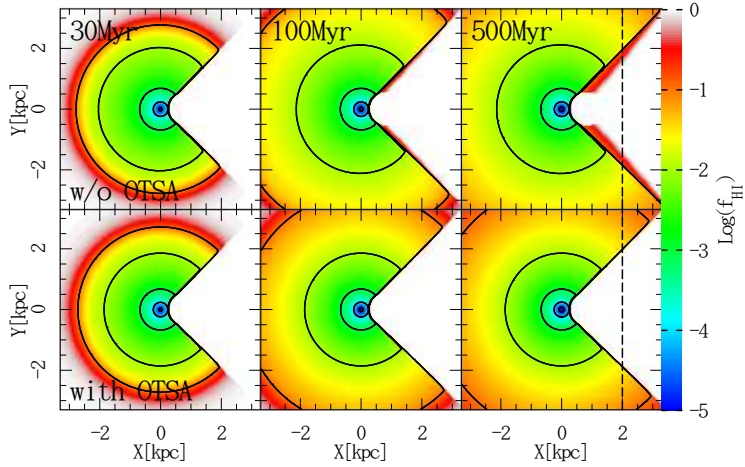}
 \caption{Test-2: Maps of neutral fraction of hydrogen in the mid-plane
 of the simulation box at $t=30{\,\rm Myr}$, $100{\,\rm Myr}$ and
 $500{\,\rm Myr}$. The lower and upper panels show the results with and
 without the on-the-spot approximation (OTSA), respectively. Dashed
 vertical lines indicate the location along which the profiles of
 temperature and neutral fractions are presented in
 Figure~\ref{fig:test-2-section}. \label{fig:test-2-map}}
\end{figure}

In Figure~\ref{fig:test-2-map}, maps of the neutral fraction of hydrogen
in the mid-plane of the simulation volume at $t=30\,{\rm Myr}$, 100Myr
and 500Myr are shown. One can see that the ionizing photons are strongly
absorbed by the dense gas clump and conical shadows are created behind
the gas clump in the both runs with and without the effect of
recombination radiation. In the run without the on-the-spot approximation
(upper panels of Figure~\ref{fig:test-2-map}), the recombination photons
emitted by the ionized gas gradually ionize the neutral gas behind the
dense gas clump.  On the other hand, in the run with the on-the-spot
approximation, the boundaries of neutral and ionized regions are kept
distinct because of the lack of recombination photons.

\begin{figure}
 \centering \includegraphics[width=10cm]{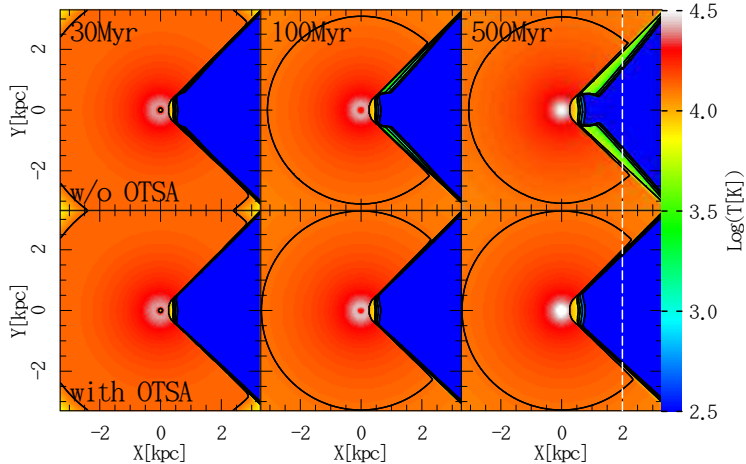}
 \caption{Test-2: Same as Figure~\ref{fig:test-2-map} but shows the gas
 temperature maps in the mid-plane of the simulation box. Dashed
 vertical lines indicate the location along which the profiles of
 temperature and neutral fractions are presented in
 Figure~\ref{fig:test-2-section}.\label{fig:test-2-t}}
\end{figure}

\begin{figure}
 \centering
 \includegraphics[width=10cm]{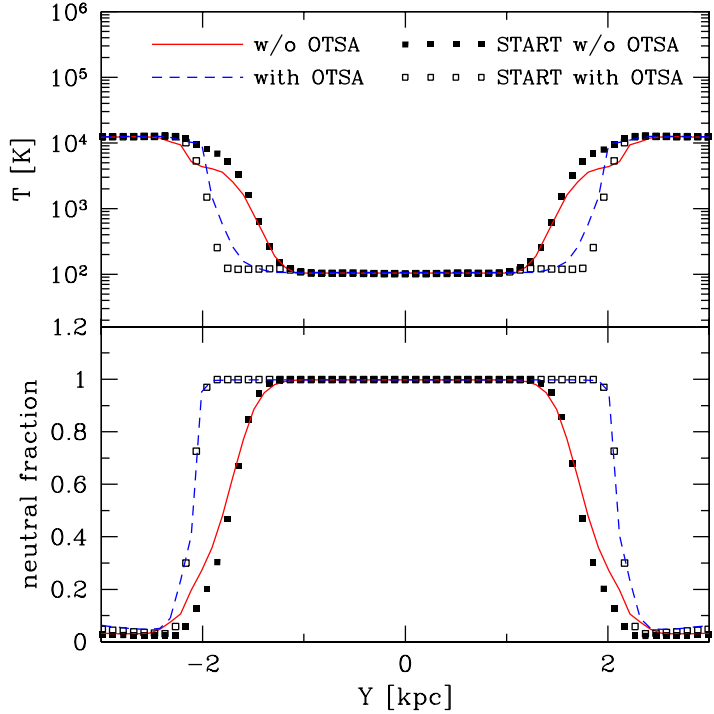}

 \caption{Profiles of temperatures and hydrogen neutral fractions along
 the dashed lines shown in Figures~\ref{fig:test-2-map} and
 \ref{fig:test-2-t} with and without the OTSA. Results in
 \citet{hasegawa2010} are also shown for
 comparison.\label{fig:test-2-section}}
\end{figure}

This test is identical to Test-6 in \citet{hasegawa2010} calculated with
the START code. In the START code, the RT is solved with a ray-tracing
scheme based on the SPH technique, and the transfer of diffuse
recombination radiation can be handled by allowing each SPH paricle to
radiate recombination photons.  Figure~\ref{fig:test-2-section} shows
profiles of gas temperature and hydrogen neutral fraction along the
lines across the conical shadow shown in Figures~\ref{fig:test-2-map}
and \ref{fig:test-2-t} as well as the results obtained with the START
code. One can see that both the results with and without the OTSA are in
fairly good agreement with each other, which supports the validity of
our scheme for diffuse radiation transfer.

\begin{figure}
 \centering \includegraphics[width=10cm]{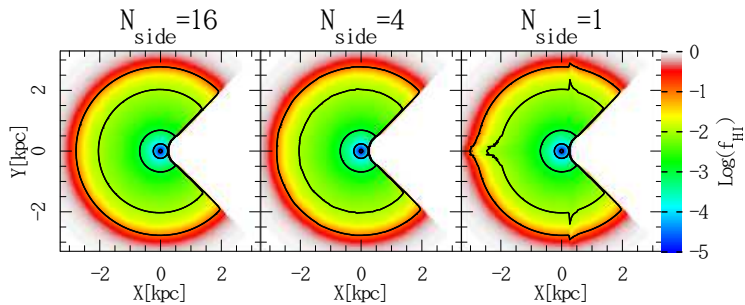}
 \caption{Test-2: Maps of neutral fraction of hydrogen in the mid-plane
 of the simulation box at $t=30{\,\rm Myr}$ for different angular
 resolution parameter, $N_{\rm side}=16$, 4 and 1 from left to right. \label{fig:test-2-resolution}}
\end{figure}

In this test, we also perform runs with various angular resolution
parameter $N_{\rm side}$ in the RT calculation of diffuse radiation to
see the effect of angular resolution.
Figure~\ref{fig:test-2-resolution} shows maps of neutral fraction of
hydrogen in the mid-plane of the simulation box at $t=30\,{\rm Myr}$
with angular resolution parameter $N_{\rm side}$ of 16, 4 and 1. The
results with $N_{\rm side}=16$ and $4$ are in good agreement with one
with $N_{\rm side}=8$ in Figure~\ref{fig:test-2-map}, indicating that
the angular resolution with $N_{\rm side}=4$ is sufficient for the
current RT calculations. The results with $N_{\rm side}=1$, however,
have spurious features in the map of neutral fraction. These numerical
artifacts can be ascribed to the low angular resolution of light-rays by
the comparison of the mean free path of the diffuse recombination
photons and $N_{\rm side}\Delta H$. As described in
\S~\ref{sub:angular_resolution}, the mean free path of the diffuse
photons should be sufficiently smaller than $N_{\rm side}\Delta H$ to
compute the RT of diffuse photons accurately. For the recombination
photons emitted by ionized hydrogens in the current setup, the mean free
path in the neutral ambient gas is estimated as
\begin{equation}
 \lambda_{\rm mfp} = \frac{1}{n_{\rm HI}\sigma_{\rm HI}(\nu_0)}=51.4
  \left(\frac{n_{\rm HI}}{10^{-3} {\rm cm}^{-3}}\right)^{-1} {\rm pc},
\end{equation}
and the mesh spacing is $\Delta H=6.6{\,\rm kpc}/128=51.5{\,\rm pc}$.
Thus, it is quite natural to have strong numerical artifacts in the
results with $N_{\rm side}=1$, because the mean free path is almost
equal to $N_{\rm side}\Delta H$, and the condition for the accurate RT
calculation ($\lambda_{\rm mfp} \ll N_{\rm side}\Delta H$) is not
satisfied.

\subsection{Test-3 : Ionization front trapping and shadowing by a dense clump}

The third test computes the transfer of ionizing radiation incident to a
face of the rectangular simulation box and the propagation of ionized
region into a spherical dense clump. This test is indentical to the
Test-3 in \citet{iliev2006}. The size of the simuation box is $6.6$ kpc,
and hydrogen number density and initial temperature are set to $n_{\rm
H}=2\times 10^{-4}$ cm$^{-3}$ and $T=8000$ K, except that a spherical
dense clump with a radius of 0.8 kpc located at 1.7 kpc apart from the
center of the simulation volume has a uniform hydrogen number density of
$n_{\rm H,c}=200n_{\rm H}=0.04$ cm$^{-3}$ and a temperature of $T_{\rm
c}=40$ K. The ionizing radiation has the blackbody spectrum with a
tempearature of $T=10^5 {\rm K}$ and constant ionizing photon flux of
$F=10^6$ s$^{-1}$ cm$^{-2}$ at a boundary of the simulation box.

\begin{figure}[htbp]
 \centering \includegraphics[width=10cm]{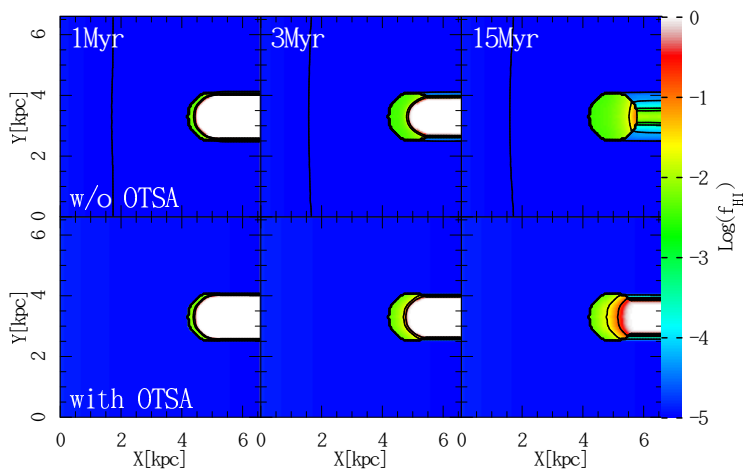} 

\caption{Test-3: Maps of neutral fraction of hydrogen in the mid-plane
 of the simulation box at $t=1$ Myr, 3 Myr and 15 Myr. The lower and
 upper panels show the results with and without the on-the-spot
 approximation, respectively.\label{fig:test-3-fmap}}
\end{figure}
\begin{figure}[htbp]
 \centering \includegraphics[width=10cm]{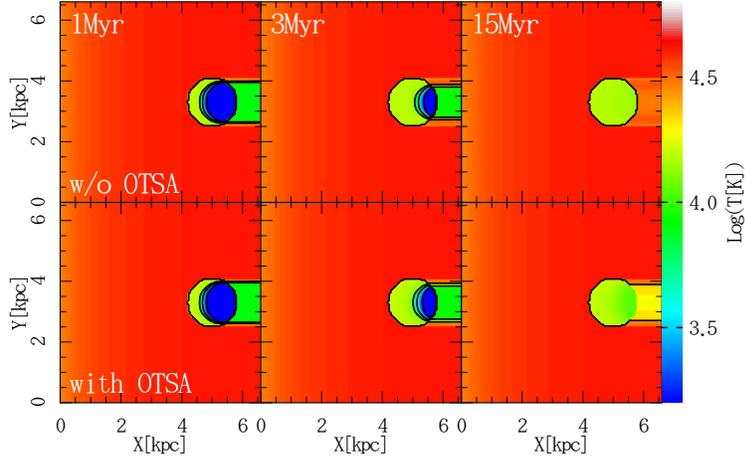} 
 
 \caption{Test-3: Same as Figure~\ref{fig:test-3-fmap} but shows the gas
 temperature maps in the mid-plane of the simulation
 box.\label{fig:test-3-tmap}}
\end{figure}

\begin{figure}[htbp]
 \centering 
 \includegraphics[width=12cm]{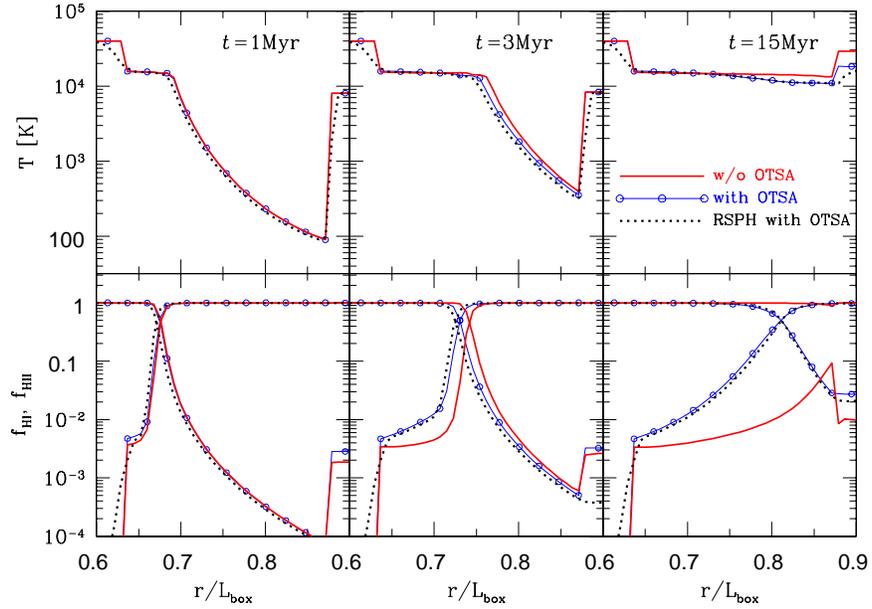}

 \caption{Test-3: Profiles of hydrogen neutral and ionized fractions and
 gas temperature along the axis of symmetry at $t=1$ Myr, 3 Myr and 15
 Myr. Solid lines with and without circles indicate the results with and
 without the on-the-spot approximation, respectively. Dotted lines shows
 the results with RSPH code~\citep{susa2006} presented in
 \citet{iliev2006} with the on-the-spot approxiamtion. \label{fig:test-3-prof}}
\end{figure}

Figures~\ref{fig:test-3-fmap} and \ref{fig:test-3-tmap} shows the maps
of hydrogen neutral fraction and gas temperature in the mid-plane of the
simulation volume at $t=1$ Myr, 3 Myr and 15 Myr from left to right,
where the ionizing photons enter from the left boundary of the
figures. We show the results with and without the on-the-spot
approximation in the lower and upper panels, respectively.

At $t=1$ Myr, the ionization front enters the spherical clump and a
cylindrical shadow is formed behind the clump. At $t=3$ Myr and 15 Myr,
the spherical clump is slightly ionized and the boundary of the shadow
is ionized and photo-heated by the hard photons which penetrate the edge
of the clump. These overall ionization and temperature structures with
the on-the-spot approximation are consistent with the ones presented in
\citet{iliev2006}. The effect of the recombination radiation is clearly
seen in the results at 15 Myr, in which the cylindrical shadow is
significantly ionized and heated by the recombination photons emitted at
the ambient ionized region.

Figure~\ref{fig:test-3-prof} shows the profiles of ionized fraction and
gas temperature along the axis of symmetry at $t=1$ Myr, 3 Myr and 15
Myr, where we also plot the results of the \verb|RSPH| code
\citep{susa2006} which are computed with the on-the-spot approximation
and presented in \citet{iliev2006}.  Our results with the on-the-spot
approximation are consistent with the ones computed with the \verb|RSPH|
code in \citet{iliev2006}. The effect of the recombination radiation is
siginificant at $t=3$ Myr and 15 Myr in the ionized fraction profiles,
in which the recombination photons accelerate the propagation of the
ionization front in the run without the on-the-spot approximation.

%

%

%

%

\section{Performance}
In this section, we show the performance of our RT calculations of
diffuse radiation. The code for the transfer of diffuse radiation is
designed so that it can be run both on multi-core CPUs and GPUs produced
by NVIDIA. The performance is measured on the HA-PACS system installed
in Center for Computational Sciences, University of Tsukuba. Each
computational node of the HA-PACS system consists of two sockets of 2.6
GHz Intel Xeon processor E5-2670 with eight cores based on the
Sandy-Bridge microarchitecture and four GPU boards of NVIDIA Tesla
M2090, each of which is connected to the CPU sockets through PCI Express
Gen2 $\times$ 16 link. Thus, a single computational node provides 2.99
Tflops (0.33 Tflops by CPUs and 2.66 Tflops by GPUs) of computing
capability in double precision.

\begin{figure}
 \centering\includegraphics[width=7cm]{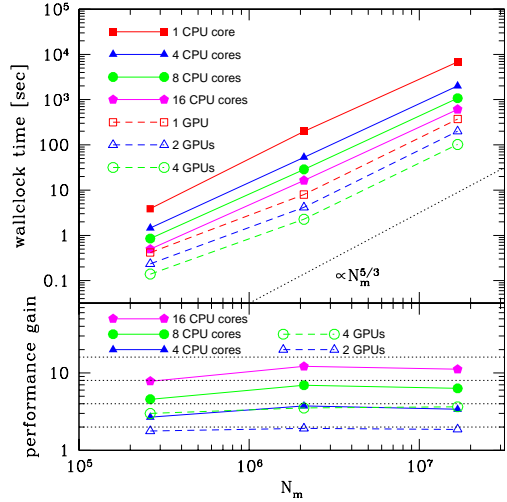}
 \caption{Wallclock times of diffuse RT calculation with various numbers
 of CPU cores and GPU boards for $N_{\rm m}=64^3$, $128^3$ and $256^3$
 are shown in the upper panel. A dotted line indicate the dependence of
 computational cost on a number of mesh grids, $\propto N_{\rm
 m}^{5/3}$. In the lower panel, we present the performance gains of
 diffuse RT calculation with multiple CPU cores and GPU boards relative
 to the performance with a single CPU core and GPU board,
 respectively. Horizontal dotted lines indicates the performance gains
 of 2, 4, 8 and 16 from bottom to top. \label{fig:time_single}}
\end{figure}

The upper panel of Figure~\ref{fig:time_single} shows wallclock time for
a iteration of the diffuse RT calculation on a single node with various
numbers of CPU cores and GPU boards. The wallclock times are measured
for $N_{\rm m}=64^3$, $128^3$ and $256^3$. The angular resolution
parameter $N_{\rm side}$ is set to $N_{\rm side} = N_{\rm m}^{1/3}/16$
so that $N_{\rm side}\Delta H$ is kept constant. Note that the wallclock
times are nearly proportional to $N_{\rm m}^{5/3}$ as theoretically
expected. The lower panel of Figure~\ref{fig:time_single} shows the
performance gain of the diffuse RT calculation with multiple CPU cores
and GPU boards relative to the performance with a single CPU core and a
single GPU board, respectively. Use of the multiple CPU cores and
multiple GPU boards provides the efficient performance gains nearly
proportional to the adopted numbers of CPU cores and GPU boards for
$N_{\rm m}=128^3$ and $256^3$ except for the fact that those with 16 CPU
cores (2 CPU sockets) is not very impressive even for $N_{\rm m}=256^3$
because of the relatively slow memory access across the CPU sockets. On
the other hand, the performance gain for $N_{\rm m}=64^3$ is somewhat
degraded because of the overheads for invoking the multiple threads and
communication overhead for data exchange between CPUs and GPUs. The
performance with the aid of four GPU boards is nearly 7 times better
than that with 16 CPU cores for $256^3$ mesh grids, while it is only 3.5
times better for $64^3$ mesh grids due to the communication overhead
between CPUs and GPUs.

\begin{figure}[tbp]
 \centering \includegraphics[width=7cm]{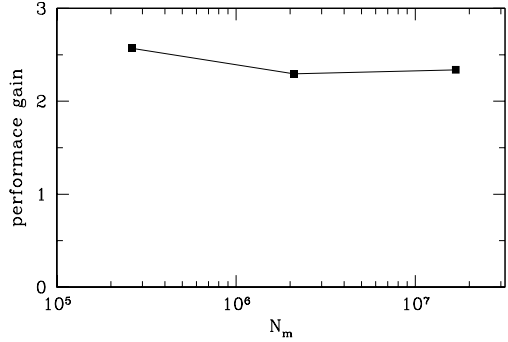}
 \caption{Performance gains obtained by the use of ray grouping for
 $N_{\rm m}=64^3$, $128^3$ and $256^3$ on GPUs. The performances are
 measured on a single computational node and four
 GPUs. \label{fig:time_atomic}}
\end{figure}

We compare the performance of our diffuse RT calculations with and
without the ray grouping technique on GPUs. In the implementation
without the ray grouping technique, we utilize the atomic operation
provided by the \verb|CUDA| programming platform in computing the
averaged radiation intensity
(equation~(\ref{eq:averaged_intensity})). Figure~\ref{fig:time_atomic}
shows the performance gains obtained by the use of the ray grouping
technique, where the individual performance is measured with a single
computational node and four GPUs. One can see that the use of the ray
grouping technique significantly improves the performance of diffuse RT
more than by a factor of two irrespective of the number of mesh grids.

\begin{figure}
 \centering \includegraphics[width=7cm]{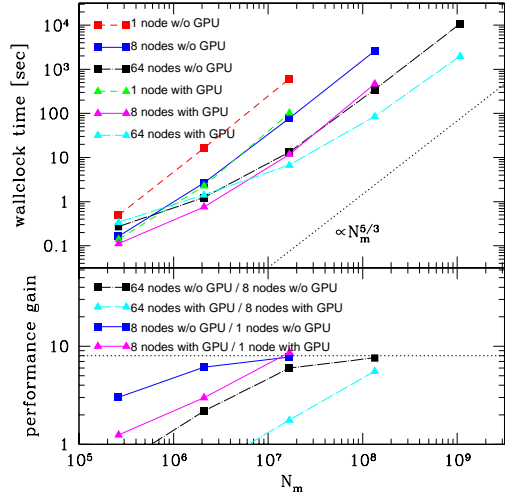}
 \caption{Wallclock times for diffuse RT calculation with various number
 of mesh grids computed with 1, 8 and 64 nodes are shown in the upper
 panel. The results with and without the use of four GPU boards are
 depicted. The dashed line indicates an analytic scaling of
 computational cost for the RT calculation of diffuse radiation,
 $N_{\rm m}^{5/3}$. \label{fig:time_multi}}
\end{figure}

\begin{figure}[htbp]
 \centering \includegraphics[width=7cm]{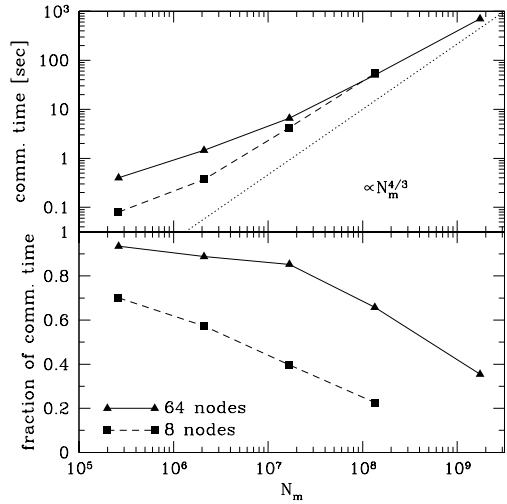}

 \caption{Upper panel: Wallclock times for MPI communication in diffuse
 radiation transfer calculations for various number of mesh grids with 8
 and 64 nodes. A scaling relation of $N_{\rm m}^{4/3}$ is shown in a
 dashed line. Lower panel: fractions of wallclock times for MPI
 communication relative to total wallclock time elapsed in the
 calculations of diffuse radiation transfer in the runs with the use of
 GPUs.\label{fig:time_comm}}
\end{figure}

The upper panel of Figure~\ref{fig:time_multi} shows the wallclock time
of diffuse RT calculation performed on a single and multiple
computational nodes with and without GPU boards, where we invoke one MPI
process on each computational node. In the runs without the use of GPUs,
each MPI process invokes 16 \verb|OpenMP| threads, while in the runs
with the aid of GPUs, we utilize four GPU boards on each computational
node. We measure the wallclock time consumed for a single iteration of
diffuse RT calculation for $64^3$--$1024^3$ mesh grids on 1, 8, and 64
computational nodes. The lower panel depicts the performance gain of the
runs with 8 and 64 computational nodes relative to those with 1 and 8
computational nodes, respectively, where the ideal performance gain of 8
is shown by a dotted line. As for the runs without the use of GPUs, the
parallel efficiency is reasonable when $N_{\rm m}/N_{\rm node} \ge
64^3$, where $N_{\rm node}$ is the number of computational nodes in
use. For a given number of computational nodes, the runs with the use of
GPUs have poorer performance gains than those without it, mainly
beacause the computational times in the runs with GPUs are significantly
shorter than those withtout GPUs, and the MPI data communication, as
well as the commnication between CPUs and GPUs, gets more salient. Such
communication overhead is proportional to the number of light-rays
getting through the surface of the decomposed computational domains,
$\propto N_{\rm m}^{2/3}N_{\rm d}\propto N_{\rm
m}^{4/3}$. Figure~\ref{fig:time_comm} shows that the time consumed by
the MPI communication is nearly proportional to $N_{\rm m}^{4/3}$, and
that it occupies a significant fraction of the total wallclock time for
a small $N_{\rm m}$. This scaling with respect to $N_{\rm m}$ has weaker
dependence on $N_{\rm m}$ than the computational costs, $\propto N_{\rm
m}^{5/3}$. Therefore, the overhead can be concealed for a sufficient
number of mesh grids, and we have better parallel efficiency for a
larger $N_{\rm m}/N_{\rm node}$.

\section{Summary \& Discussion}

In this paper, we present a new implementation of the RT calculation of
diffuse radiation field on three-dimensional mesh grids, which is
suitable to be run on recent processors with highly-parallel
architecture such as multi-core CPUs and GPUs. The code is designed to
be run on both of ordinary multi-core CPUs and GPUs produced by NVIDIA
by utilizing the
\verb|OpenMP| application programming interface and the \verb|CUDA|
programming platform, respectively.

Since our RT calculation is based on the ray-tracing scheme, the RT
calculation itself can be carried out concurrently by assigning the RT
calculation along each light-ray to individual software threads. To
avoid the atomic operations in computing the averaged intensity
(equation~(\ref{eq:path_averaged_intensity})) which can potentially
degrade the efficiency of the thread parallelization, we devise a new
scheme of the RT calculations in which a set of parallel light-rays are
split into 4 groups so that parallel light-rays in each group do not get
through any mesh grids more than once.  As well as the thread
parallelization inside processors or computational nodes, we also
parallelize our code on a multi-node system using the MWF scheme
developed by \citet{nakamoto2001}.

We perform several test simulations where the transfer of photo-ionizing
radiation emitted by a point radiating source and recombination
radiation from ionized regions as diffuse radiation are solved. We
verify the validity of our RT calculation of the diffuse radiation by
comparing our results with the effect of recombination radiation and the
ones with other two independent codes, the one-dimensional spherical
code by \citet{kitayama2004} and START code by \citet{hasegawa2010}.  We
also clarify the condition of the required angular resolution in our
diffuse radiation transfer scheme based on the mean free path of the
diffuse photons and the mesh spacing.

We show good parallel efficiency of our implementation for intra- and
inter-node parallelizations. As for the intra-node parallelization, the
performance scales well with the number of CPU cores and GPU boards in
use, except for the one in the case that multiple CPU sockets are used
as a single shared-memory system. The scalability of the inter-node
parallelization with the MWF scheme is also measured for $64^3$ to
$1024^3$ mesh grids on up to 64 computational nodes and it is found that
the inter-node parallelization is efficient when we have a sufficient
number of mesh grids per node, $N_{\rm m}/N_{\rm node}\ge 128^3$ and
$N_{\rm m}/N_{\rm node}\ge 64^3$ for the runs with and without GPUs,
respectively. The ray-grouping technique described in
\ref{ss:ray_grounping} is effective and significantly improves the
performance of our RT calculations by a factor of more than two, at
least on GPUs (NVIDIA Tesla M2090).

With our implementation presented in this paper, we are able to perform
the diffuse RT calculations in a reasonable wallclock time comparable to
that of other physical processes such as hydrodynamical
calculations. This means that the calculations of the diffuse radiation
transfer can be coupled with hydrodynamic simulations and we are able to
conduct radiation hydrodynamical simulations with the effect of diffuse
radiation transfer as well as the radiation transfer from point
radiating sources in three-dimensional mesh grids. Currently, we are
developing such a radiation hydrodynamic code and, based on this, we
will address astrophysical problems in which diffuse radiation transfer
plays important roles.

It should be noted that, though we present the implementations and the
performance on the multi-core CPUs and GPUs produced by NVIDIA, our
approaches presented in this paper can be readily applied to other
processors with similar architecture, such as the Intel Xeon-Phi
processor or GPUs by other vendors. In addition, our approach can be
easily extended to adaptively refined mesh grids using the prescription
described in \citet{razoumov2005}, although we present the
implementation for uniform mesh grids in this paper.

\bigskip

We would like to thank Tetsu Kitayama for providing us with his
radiation transfer code. We are also grateful to the anonymous referee
for helpful comments. Numerical simulations in this work have been
carried out on the HA-PACS supercomputer system under the
``Interdisciplinary Computational Science Program'' in the Center for
Computational Sciences, University of Tsukuba. This work was partially
supported by Japan Society for the Promotion of Science (JSPS)
Grant-in-Aid for Scientific Research (S) (20224002). TO acknowledges the
financial support of JSPS Grant-in-Aid for Young Scientists (B:
24740112). KH acknowledges the support of MEXT SPIRE Field 5 and JICFuS
and the financial support of JSPS Grain-in-Aid for Young Scientists (B:
24740114).

\appendix
\section{Photon-conserving estimation of photo-heating and radiative cooling rates\label{app:photon_conserving}}

We describe the photon-conserving evaluation of photo-ionization and
photo-heating rates of a mesh grid contributed by a point radiating
source, where the inner and outer intersections of a light-ray from the
point raidating source and the mesh grid are located at $r_{\rm in}$ and
$r_{\rm out}$ from the point radiating source. We consider a imaginary
spherical shell centered at the point radiating source with inner and
outer radii of $r_{\rm in}$ and $r_{\rm out}$, respectively. The
incoming photon number per unit time $\dot{N}_{{\rm in},\nu}$ is given
by
\begin{equation}
\dot{N}_{{\rm in}, \nu} = \frac{L_{\nu} \exp(-\tau_{\nu})}{h\nu},
\end{equation}
where $L_{\rm \nu}$ is the luminosity density at a frequency $\nu$, and
$\tau_{\nu}$ is the optical depth between the point source and the inner
side of the shell. The outgoing photon number per unit time from the
outer side of the shell is written as
\begin{equation}
\dot{N}_{{\rm out}, \nu} = \frac{L_{\nu} \exp(-(\tau_{\nu} +\Delta \tau_{\nu}))}
{h\nu} ,
\end{equation}
where $\Delta \tau_\nu$ is the radial optical depth of the shell.  Then,
the number of absorped photons per unit time $\dot{N}_{\rm abs}$ is
given by
\begin{equation}
 \dot{N}_{{\rm abs}, \nu} = \dot{N}_{{\rm out},\nu} - \dot{N}_{{\rm in},\nu} = \frac{L_{\nu} \exp(-\tau_{\nu})}{h\nu} \left[1-\exp(-\Delta \tau_{\nu}) \right] .
\end{equation}
When we consider multiple chemical components, the absoption rate of the
$i$-th species is rewritten as
\begin{equation}
 \dot{N}^i_{{\rm abs}, \nu} = \frac{\Delta \tau^i_\nu}{\Delta \tau_{\nu}}\frac{L_{\nu} \exp(-\tau_{\nu})}{h\nu} \left[1-\exp(-\Delta \tau_{\nu}) \right],
\end{equation}
where $\Delta \tau^i_\nu$ is a optical depth contributed by the $i$-th
component, and $\Delta \tau_\nu = \sum_i \Delta \tau^i_\nu$ is the total
optical depth. Since $\dot{N}^i_{{\rm abs},\nu}$ is equal to the number
of ionization of the $i$-th species, the photo-ionization rate of the
$i$-th component can be written as
\begin{equation}
 \Gamma_{i,\gamma} = \frac{1}{N_{i}} \int^{\infty}_{\nu_{i}}
  \dot{N}^{i}_{{\rm abs},\nu} d\nu ,
\label{eq:pc_gamma}
\end{equation}
where $\nu_i$ is the threshold frequecy of the $i$-th species and $N_i$
is the number of $i$-th species in the shell. The photo-heating rate is
similarly calculated in terms of $\dot{N}_{{\rm abs},\nu}$ as
\begin{equation}
 {\cal H}_{i,\gamma} = \frac{1}{N_i}\int_{\nu_i}^\infty \dot{N}_{{\rm abs},\nu} (h\nu-h\nu_i) d\nu.
\end{equation}

\section{Ionization Balance\label{app:reaction}}

The time evolution of the number density of the $i$-th chemical species
can be schematically described by
\begin{equation}
 \label{eq:app_chem}
 \frac{dn_i}{dt} = C_i(T,n_j)-D_i(T,n_j)n_i, 
\end{equation}
where $C_i(T,n_j)$ is the collective production rate of the $i$-th
species and $D_i(T, n_j)n_i$ is the destruction rate of the $i$-the
species. For example, in the case of atomic hydrogen, $C_{\rm HI}$ and
$D_{\rm HI}$ is given by
\begin{eqnarray}
 C_{\rm HI} &=& \alpha_{\rm HII}\,n_{\rm e}n_{\rm HII}\\
 D_{\rm HI} &=& \Gamma_{\rm HI} n_{\rm e} + \Gamma_{\rm HI,\gamma}
\end{eqnarray}
where $\alpha_{\rm HII}(T)$ is the radiative recombination rate of HII,
$\Gamma_{\rm HI}$ is the collisional ionization rate and $\Gamma_{\rm
HI,\gamma} = \sum_\alpha\Gamma^\alpha_{\rm HI,\gamma} + \Gamma_{\rm
HI,\gamma}^{\rm diff}$ is the photoionization rate of HI.

These equations are numerically solved using the backward difference
formula (BDF) \citep{anninos1997,yoshikawa2006}, in which the number
densities of the $i$-th chemical species at a time $t+\Delta t$,
$n_i^{t+\Delta t}$, is computed as
\begin{equation}
 n_i^{t+\Delta t} = \frac{C_i\Delta t+n_i^t}{1+D_i\Delta t},
\end{equation}
where, $C_i$ and $D_i$ are estimated with the number densities of each
species at the advanced time, $n_j^{t+\Delta t}$. However, the number
densities in the advanced time step are not available for all the
chemical species in evaluating $C_i$ due to the intrinsic non-linearity
of equation~\ref{eq:app_chem}. Thus, we sequentially update the number
densities of each chemical species in the increasing order of ionization
levels rather than updating all the species simultaneously. It is
confirmed that this scheme is stable and accurate \citep{anninos1997,
yoshikawa2006}.

\section{Photo-heating and radiative cooling\label{app:heating_cooling}}

The specific energy change for each mesh by the photo-heating and
radiative cooling is followed by the energy equation
\begin{equation}
 \label{eq:app_energy}
 \frac{du}{dt} = \frac{\mathcal{H}-\mathcal{C}}{\rho},
\end{equation}
where $u$ is the specific internal energy and $\mathcal{H}$ and
$\mathcal{C}$ are the photo-heating and cooling rate, respectively and
$\mathcal{H}$ is given by
\begin{equation}
 \mathcal{H} = \sum_i n_i \left(\sum_\alpha \mathcal{H}_i^\alpha + \mathcal{H}_i^{\rm diff}\right).
\end{equation}
The specific internal energy for each mesh is updated implicitly by
solving the equation
\begin{equation}
 u^{t+\Delta t} = u^t + \frac{\mathcal{H}^{t+\Delta t}
  -\mathcal{C}^{t+\Delta t}}{\rho^t}\Delta t
\end{equation}
for $u^{t+\Delta t}$, where the photo-heating $\mathcal{H}^{t+\Delta t}
= \mathcal{H}(n^{t+\Delta t})$ and cooling rates $\mathcal{C}^{t+\Delta
t} = \mathcal{C}(n^{t+\Delta t}, u^{t+\Delta t})$ are evaluated at the
advanced time $t+\Delta t$.

\section{Timestep constraints\label{app:timestep}}

Since we solve the static RT equation (\ref{eq:rad_tr1}), equations
(\ref{eq:app_chem}) for chemical reactions and (\ref{eq:app_energy}) for
photo-heating and radiative cooling have to be solved iteratively until
the electron number density and specific internal energy in each mesh
grid converges: $|n_{\rm e}^{(i)}-n_{\rm e}^{(i-1)}|<\epsilon n_{\rm
e}^{(i)}$ and $|u^{(i)}-u^{(i-1)}|<\epsilon u^{(i)}$, where $\epsilon$
is set to $10^{-3}$, and $n_{\rm e}^{(i)}$ and $u^{(i)}$ indicates the
specific internal energy and the electron number density after the
$i$-th iteration, respectively.

The timestep in solving chemical reactions and energy equation, $\Delta
t_{\rm chem}$, is set to
\begin{equation}
 \Delta t_{\rm chem} = \epsilon_{\rm e}\left|\frac{n_{\rm e}}{\dot{n}_{\rm
			   e}}\right|+\epsilon_{\rm HI} \left|\frac{n_{\rm
			   HI}}{\dot{n}_{\rm HI}}\right|,
\end{equation}
where the second term on the right hand side prevents the timestep from
getting prohibitively short in the case that the gas is almost
neutral, and $\epsilon_{\rm e}$ and $\epsilon_{\rm HI}$ are set to 0.2
and 0.002, respectively. 

The timestep, $\Delta t$, with which we update the radiation field can
be larger than the chemical timestep, $\Delta t_{\rm chem}$ by
subcycling the rate and energy equations (\ref{eq:app_chem}) and
(\ref{eq:app_energy}). Throughout in this paper, the timestep for the RT
calculation is set to
\begin{equation}
 \Delta t = F \min_i \Delta t_{\rm chem,\it i}, 
\end{equation}
where $\Delta t_{\rm chem,\it i}$ is the chemical timestep for the
$i$-th mesh grid, and we typically set $F=1\sim10$ so that the radiation
field successfully converges with a reasonable number of iterations.


\begin{thebibliography}{}

\bibitem[Abel et~al.(1997)]{abel1997} Abel, T., Anninos, P., Zhang, Y.,
Norman, M.~L., 1997, New Astronomy, 2, 181

\bibitem[Abel, Norman \& Madau(1999)]{abel1999}
Abel, T., Norman,M.~L., Madau, P.,  1999, \apj, 523, 66

\bibitem[Anninos et~al.(1997)]{anninos1997}
Anninos, P., Zhang, Y., Abel, T., Norman, M.~L.,  1997, New Astronomy, 2, 209

\bibitem[Cen(1992)]{cen1992}
Cen, R.,  1992, \apjs, 78, 341

\bibitem[Ciardi et al.(2001)]{ciardi2001} Ciardi, B., Ferrara, A., 
Marri, S., \& Raimondo, G.\ 2001, \mnras, 324, 381

\bibitem[Dopita et~al.(2011)]{dopita2011}
Dopita, M.~A.,  Krauss, L.~M.,  Sutherland, R.~S.,  Kobayashi, C.,
		Lineweaver, C.~H.,  2011, \apss, 335, 345

\bibitem[Gnedin \& Abel(2001)]{gnedin2001}
Gnedin, N.~Y.,  Abel, T.,  2001, New Astronomy, 6, 437

\bibitem[Gonz{\'a}lez, Audit \& Huynh(2007)]{gonzalez2007}
Gonz{\'a}lez, M.,  Audit, E.,    Huynh, P.,  2007, \aap, 464, 429

\bibitem[G{\'o}rski et~al.(2005)]{gorski2005}
G{\'o}rski, K.~M.,  Hivon, E.,  Banday, A.~J.,  Wandelt, B.~D.,  Hansen,
  F.~K.,  Reinecke, M.,   Bartelmann, M.,  2005, \apj, 622, 759

\bibitem[Hasegawa \& Umemura(2010)]{hasegawa2010}
Hasegawa, K., Umemura, M.,  2010, \mnras, 407, 2632

\bibitem[Hui \& Gnedin(1997)]{hui1997}
Hui, L., Gnedin, N.~Y.,  1997, \mnras, 292, 27

\bibitem[Hummer(1994)]{hummer1994}
Hummer, D.~G.,  1994, \mnras, 268, 109

\bibitem[Hummer \& Storey(1998)]{hummer1998}
Hummer, D.~G.,  Storey, P.~J.,  1998, \mnras, 297, 1073

\bibitem[Ikeuchi \& Ostriker(1986)]{ikeuchi1986}
Ikeuchi, S., Ostriker, J.~P.,  1986, \apj, 301, 522

\bibitem[Iliev et~al.(2006)]{iliev2006}
Iliev, I.~T.  et~al., 2006, \mnras, 371, 1057

\bibitem[Iliev et~al.(2009)]{iliev2009}
Iliev, I.~T.  et~al., 2009, \mnras, 400, 1283

\bibitem[Inoue(2010)]{inoue2010} 
Inoue, A.~K.\ 2010, \mnras, 401, 1325

\bibitem[Janev et~al.(1987)]{janev1987}
Janev R.~K.,  Langer W.~D.,  Post~Jr D.~E.,    Evans~Jr K.,  1987, in
		Janev R.K., Lnger W.D., Evans, K., eds, Elementary
		Processes in Hydrogen-Helium Plasmas -- Cross-section
		and Reaction Rate Coefficients. Springer-Verlag , Berlin

\bibitem[Kanno et~al.(2013)]{kanno2013}
Kanno, Y., Harada, T., \& Hanawa, T. 2013, \pasj, 65, 72


\bibitem[Kitayama et~al.(2004)]{kitayama2004}
Kitayama, T., Yoshida, N., Susa, H.,  Umemura, M.,  2004, \apj, 613,
  631

\bibitem[Kunasz \& Auer(1988)]{kunasz1988}
Kunasz, P., Auer, L.~H.,  1988, J. Quant. Spectr. Radiative Transfer, 39, 67

\bibitem[Miralda-Escud{\'e}(2003)]{miralda-escude2003}
Miralda-Escud{\'e}, J.,  2003, \apj, 597, 66

\bibitem[Nakamoto et~al.(2001)]{nakamoto2001}
Nakamoto, T.,  Umemura, M., Susa, H.,  2001, \mnras, 321, 593

\bibitem[Nakamoto et al.(2001b)]{nakamoto2001b} Nakamoto, T., Umemura,
		M., \& Susa, H.\ 2001, in ASP Conf. Ser. 222, The
		Physics of Galaxy Formation, ed. M. Umemura \& H. Susa
		(San Francisco: ASP), 109

\bibitem[Okamoto et~al.(2012)]{okamoto2012}
Okamoto, T.,  Yoshikawa, K., Umemura, M.,  2012, \mnras, 419, 2855

\bibitem[Okamoto et al.(2014)]{okamoto2014} Okamoto, T., Shimizu,
I., \& Yoshida, N.\ 2014, \pasj, 66, 70 

\bibitem[Osterbrock(2006)]{agnagn}
Osterbrock, D.~E.,  2006, Astrophysics Of Gaseous Nebulae And Active Galactic
  Muclei.
University Science Books

\bibitem[Pawlik \& Schaye(2011)]{pawlik2011}
Pawlik, A.~H.,  Schaye, J.,  2011, \mnras, 412, 1943

\bibitem[Rahmati et al.(2013a)]{rahmati2013a} Rahmati, A., Pawlik, 
A.~H., Rai\v{c}evi\'{c}, M., \& Schaye, J.\ 2013, \mnras, 430, 2427

\bibitem[Rahmati et al.(2013b)]{rahmati2013b} Rahmati, A., Schaye, 
J., Pawlik, A.~H., \& Rai\v{c}evi\'{c}, M.\ 2013, \mnras, 431, 2261 

\bibitem[Razoumov \& Cardall(2005)]{razoumov2005}
Razoumov, A.~O.,  Cardall, C.~Y.,  2005, \mnras, 362, 1413

\bibitem[Rijkhorst et~al.(2006)]{rijkhorst2006}
Rijkhorst, E.-J.,  Plewa, T.,  Dubey, A.,    Mellema, G.,  2006, A\&A, 452,
  907

\bibitem[Rosdahl et al.(2013)]{rosdahl2013} Rosdahl, J., Blaizot, 
J., Aubert, D., Stranex, T., \& Teyssier, R.\ 2013, \mnras, 436, 2188 

\bibitem[Skinner \& Ostriker(2013)]{skinner2013} Skinner, M.~A., \&
		Ostriker, E.~C.\ 2013, \apjs, 206, 21

\bibitem[Sokasian et~al.(2001)]{sokasian2001}
Sokasian, A.,  Abel, T.,    Hernquist, L.~E.,  2001, New Astronomy, 6, 359

\bibitem[Stone et~al.(1992)]{stone1992}
Stone, J.~M.,  Mihalas, D., Norman, M.~L.,  1992, \apjs, 80, 819

\bibitem[Susa(2006)]{susa2006}
Susa, H.,  2006, \pasj, 58, 445

\bibitem[Wise \& Abel(2011)]{wise2011}
Wise, J.~H.,  Abel, T.,  2011, \mnras, 414, 3458

\bibitem[Wyithe et al.(2011)]{wyithe2011} Wyithe, J.~S.~B., Mould, 
J., \& Loeb, A.\ 2011, \apj, 743, 173

\bibitem[Yoshikawa \& Sasaki(2006)]{yoshikawa2006}
Yoshikawa, K.,  Sasaki, S.,  2006, \pasj, 58, 641

\end{thebibliography}
\end{document}